\documentclass[aps,prl,nofootinbib,superscriptaddress,letterpaper,amsmath,amssymb]{revtex4}
\pdfoutput=1

\usepackage{graphicx}  
\usepackage{dcolumn}   
\usepackage{bbm}        
\usepackage{amsmath}
\usepackage{amsfonts}
\usepackage{amssymb}   
\usepackage{feynmp}
\usepackage{epstopdf}
\usepackage{slashed}
\usepackage{multirow}
\usepackage{color}
\usepackage{float}

\usepackage{verbatim}

\def\gsim{\lower0.5ex\hbox{$\:\buildrel >\over\sim\:$}}
\def\lsim{\lower0.5ex\hbox{$\:\buildrel <\over\sim\:$}}

\newcommand{\be}{\begin{equation}}
\newcommand{\ee}{\end{equation}}
\newcommand{\bea}{\begin{eqnarray}}
\newcommand{\eea}{\end{eqnarray}}

\newcommand{\nbox}{{\,\lower0.9pt\vbox{\hrule \hbox{\vrule height 0.2 cm
\hskip 0.2 cm \vrule height 0.2 cm}\hrule}\,}}

\def\sub#1{_{\lower.25ex\hbox{$\scriptstyle#1$}}}

\newskip\zatskip \zatskip=0pt plus0pt minus0pt
\def\matth{\mathsurround=0pt}
\def\lsim{\mathrel{\mathpalette\atversim<}}
\def\gsim{\mathrel{\mathpalette\atversim>}}
\def\sigv{\ifmmode \langle\sigma v\rangle\else $\langle\sigma v\rangle$\fi}
\newskip\zatskip \zatskip=0pt plus0pt minus0pt
\def\matth{\mathsurround=0pt}
\def\lsim{\mathrel{\mathpalette\atversim<}}
\def\gsim{\mathrel{\mathpalette\atversim>}}
\def\atversim#1#2{\lower0.7ex\vbox{\baselineskip\zatskip\lineskip\zatskip
  \lineskiplimit
  0pt\ialign{$\matth#1\hfil##\hfil$\crcr#2\crcr\sim\crcr}}}

\begin{document}

\thispagestyle{empty}
\vspace*{-3.5cm}

\vspace{0.5in}

\title{Supersoft SUSY Models and the 750 GeV Diphoton Excess, Beyond Effective Operators   }

\begin{center}
\begin{abstract}
We propose that the sbino, the scalar partner of a Dirac bino can explain the 750 GeV diphoton excess observed by the ATLAS and CMS collaborations.  We analyze a model in which the sbino couples to pairs of Standard Model (SM) gauge bosons through effective operators. Additionally, we consider a completion in which the sbino couples to pairs of gauge bosons through loops of heavy sfermions.  We find that the sbino may be given an appreciable decay width through tree level coupling in the Higgs or Higgsino sector and additionally fit the 750 GeV excess by considering gluon fusion processes with decay to diphotons.

\end{abstract}
\end{center}

\author{Linda M. Carpenter}
\affiliation{The Ohio State University, Columbus, OH}
\author{Russell Colburn}
\affiliation{The Ohio State University, Columbus, OH}

\author{Jessica Goodman}
\affiliation{The Ohio State University, Columbus, OH}

\maketitle


\section{Introduction}


Both the CMS and ATLAS collaborations have reported an excess in the diphoton resonance channel in the first stages of the LHC's 13 TeV run.  This excess seems to correspond to a new boson with mass of approximately 750 GeV.  ATLAS reports a diphoton resonance with mass 747 GeV at 3.6$\sigma$ local significance while CMS reports a diphoton resonance of mass 760 GeV and a local significance of 2.6$\sigma$ \cite{ATLAS,CMS}.  If the excess persists, it would be a smoking gun for a new sector beyond the SM.  Many beyond the SM (BSM) scenarios might accommodate such a resonance including models with exotic axion-like states, models with strong couplings, extra dimension, heavy Higgses and more \cite{Mambrini:2015wyu}-\cite{Harigaya:2015ezk}.  The production process of the new state may be variable, the diphoton resonance may be produced in association with other states, or alone \cite{ATLAS}.  Thus, production may be through gluon or quark fusion to a single new resonance, vector boson fusion, or production in association with other states.

The simplest assumption for a BSM particle candidate which decays to two photons is a scalar field that couples to SM gauge bosons through a dimension 5 operator.  This would indicate the existence of some heavy ``messenger" particles with SM charges that couple both to SM gauge bosons and to the singlet field responsible for the possible excess.  This mechanism is, in principle, exactly the same as the SM Higgs coupling to photon and gluon pairs. One guess, then, for the identity of the possible 750 GeV state is simply a heavy Higgs as would appear in a Type II two Higgs doublet model like the  Minimal Supersymmetric Standard Model (MSSM).  This possibility would be quite exciting as supersymmetry (SUSY) is the leading candidate for BSM physics.  However, it has been pointed out that a heavy Higgs in the minimal MSSM scenario fails to reproduce the observed rate of the excess.  The MSSM would have to be extended by adding multiple sets of vector like chiral superfields to enhance the signal \cite{Angelescu:2015uiz}. There are, however, alternate SUSY scenarios which would fit the excess.

We propose that a well studied supersymmetric scenario contains all of the necessary pieces to fit the excess, that of an R symmetric MSSM. In these scenarios gauginos are Dirac particles, rather than Majorana \cite{Fox:2002bu,Hall:1990hq}. The Dirac gaugino gets its mass by ``marrying" a chiral SM adjoint field.  The superpartners of these new fermion fields are complex scalar particles in the adjoint representation of their respective gauge symmetry. Thus, there exists scalars and pseudoscalars which are a color octet, a SU(2) triplet, and SM singlet fields.  The bino superpartner offers a scalar and pseudoscalar SM singlet candidate, and in this work we explore the possibility that the 750 GeV resonance is the real scalar part of the SM singlet superfield whose fermionic component marries the bino.  In keeping with R symmetric parlance we can refer to this as the sbino.

The sbino fields may couple to SM gauge bosons in a variety of ways.  First we may simply consider a set of general effective operators which couple scalar fields to SM field strength tensors.  These operators are consistent with all symmetries of the theory, and extremely similar to operators which produce the Dirac gaugino masses themselves.  We may complete these operators through loops in which the scalar fields couple to a heavy fermion or scalars which are charged under the SM gauge groups.  However, these heavy fields need not be added to the model to explain the excess. In the case of fermion these may simply be the messenger fields already necessary to create Dirac gaugino masses.  Even simpler, and the case we will study in this work, the messenger fields may be the superpartners of the fermions themselves which have Kahler potential couplings to the new chiral superfields. This means that Dirac gaugino models already contain the necessary fields, operators, and couplings to produce the diboson coupling needed to explain the LHC excess.  We will explore a simplified version of a Dirac gaugino model calculating the loop level couplings of the sbino to pairs of gluons and photons. We calculate the gluon fusion production cross section and calculate relative decay rates into gluons, photons, and light Higgs fields of the sbino.  We find that we can approximately match the large decay width and the diphoton rate at 13 TeV.

This paper is organized as follows, in Section II we review the formalism of models with Dirac gauginos and introduce couplings of the sbino to pairs of dibosons as an effective operator.  In section III we discuss the UV generation of the effective operator through Kahler potential couplings between the sbino and the SM sparticles.  In section IV we give relative branching rations of the state and compute its production through gluon fusion and decay rate to photons, and discuss current collider constraints of this scenario.  In section V we conclude.


\section{Operators in Dirac Gauginos Models}

We consider a class of SUSY models in which gaugino mass terms are Dirac as opposed to Majorana \cite{Fox:2002bu,Hall:1990hq}.
In such models, the gauginos get mass by ``marrying" chiral fields which are adjoints under SM gauge groups.
The Dirac mass is generated by the superpotential operator
\begin{equation}
W= \int d^2 \theta \frac{W^{'}_{\alpha}W^{\alpha}A}{\Lambda} \supset \frac{D' }{\Lambda}\lambda \psi_A
\label{eq:ssoft}
\end{equation}
where  W is the standard model gauge field strength tensor, A is the chiral adjoint field, and $W'$ is the field strength tensor of a hidden $U(1)$ gauge group.  The hidden sector U(1) is broken at a high scale and gets a nonzero D-term vev.  After the symmetry breaking this operator, known as the supersoft operator, contains a Dirac mass for the gaugino with mass $m_D \equiv D/\Lambda$.

This SUSY breaking can be embedded in the framework of gauge mediation \cite{Dine:1995ag}.  One may add to the theory a set of heavy messenger particles that are charged under the SM and the new U(1) gauge group.  These messengers couple the SUSY breaking sector to the gauginos and SM adjoint fields, thus communicating the SUSY breaking to the visible sector.

The new chiral multiplets, A, contain real and imaginary scalar degrees of freedom. One might assume that the real and imaginary fields have large mass of order $m_D$. However, the masses of the real and imaginary parts of the multiplets may be quite split. These masses depend greatly on the details of the messenger sector of the model \cite{Carpenter:2010as},\cite{Carpenter:2015mna,Csaki:2013fla}. Since most models have large negative mass contributions for one combination of the fields, it is quite natural to assume that one of the adjoints is much lighter than the other, and may be thus accessible to colliders in the TeV range.

Consistent with all symmetries of the theory, we may also write a set of operators extremely similar to those that yield the Dirac gaugino masses in Eq. \ref{eq:ssoft}.  These operators couple the scalar adjoint fields to the square of SM field strength tensors.  We will denote the new chiral fields as follows; $O$ denotes the SU(3) adjoint, $T$ the SU(2) adjoint, and $S$ the SM singlet. We then have a complete set of gauge invariant operators

\begin{equation}
W= \int d^{2}\theta \frac{W_{Y \alpha} W^{\alpha}_Y S}{\Lambda_1}+\frac{W_{2 \alpha} W^{\alpha}_2 T}{\Lambda_2}+\frac{W_{3 \alpha} W^{\alpha}_3 O}{\Lambda_3}
\label{eq:StoGaugeB}
\end{equation}
where $W_Y$ is the hypercharge field strength, $W_i$ is the appropriate SU(2) or SU(3) field strength tensor, and $\Lambda_i$ is the appropriate cut-off scale which may be different for each operator.  Integrating over the fermionic coordinates these operators become terms in the Lagrangian,
\be
\mathcal{L}\supset\frac{1}{\Lambda_{1}} S_r B^{\mu \nu} B_{ \mu \nu}+\frac{1}{\Lambda_{2}} T_r W^{\mu \nu} W_{ \mu \nu}+\frac{1}{\Lambda_{3}} O_r G^{\mu \nu} G_{ \mu \nu}
\ee

\noindent These operators couple the real part (as indicated by the subscript) of the scalar states to the square of the SM field strength tensors.  Similarly, we may expect to generate operators which couple the imaginary part of the scalar states to the SM field strength tensor and its dual.
\be
\mathcal{L}=\frac{1}{\Lambda_{1}} S_i B^{\mu \nu} \tilde{B}_{ \mu \nu}+\frac{1}{\Lambda_{2}} T_i W^{\mu \nu} \tilde{W}_{ \mu \nu}+\frac{1}{\Lambda_{3}} O_i G^{\mu \nu} \tilde{G}_{ \mu \nu}
\ee

The full set of such operators were discussed in reference \cite{Carpenter:2015gua}.  Interestingly, we may also write in the Lagrangian a gauge invariant term which couples pairs of gluons to the SU(2) adjoint, with Higgs fields inserted to ensure gauge invariance.  This operator provides another interesting prospect for scalar states accessible by gluon fusion.

\section{UV completions}

We expect that the operators coupling the SM field strength tensors to the chiral adjoints may be completed by considering integrating out loops of heavy ``messengers".  One possibility is simply to consider loops of the heavy messengers which generate the Dirac gaugino masses themselves.  These fields carry SM quantum numbers, the simplest messenger sector are those in which the messengers are  fundamentals and antifundamentals under SU(5), see for example \cite{Carpenter:2015mna,Csaki:2013fla}.  Though messengers may couple to the chiral adjoint through large Yukawa like couplings, we expect that the messenger mass scale is very large, thereby suppressing the operator.  Another possibility is to consider loops containing the scalar superpartners of the fermions.  

Dirac gaugino models have a unique Kahler potential coupling between the new chiral states and the standard model sfermions generated by integrating out SM D terms.  The scalar potential involving the singlet scalar contains the terms
\be
V\sim m_D (S+S^{*}) D_Y + q g_Y D_Y \tilde{f}  {\tilde{f}}^{\dagger} + \frac{1}{2}D_YD_Y
\ee \noindent
where $\tilde{f}$ are the SM sfermions, q are the hypercharges of each sfermion, and $D_Y$ is the U(1) hypercharge D-term.  Once the SM D-term is integrated out, one finds a trilinear coupling between the real component of the scalar and each pair of sfermions with non-zero hypercharge,

\be
V\sim q g_Y m_D (S+S^{*})\tilde{f}  {\tilde{f}}^{\dagger}.
\label{eq:cubic}
\ee

\begin{figure}[tbh]
\centering
\includegraphics[scale=.7]{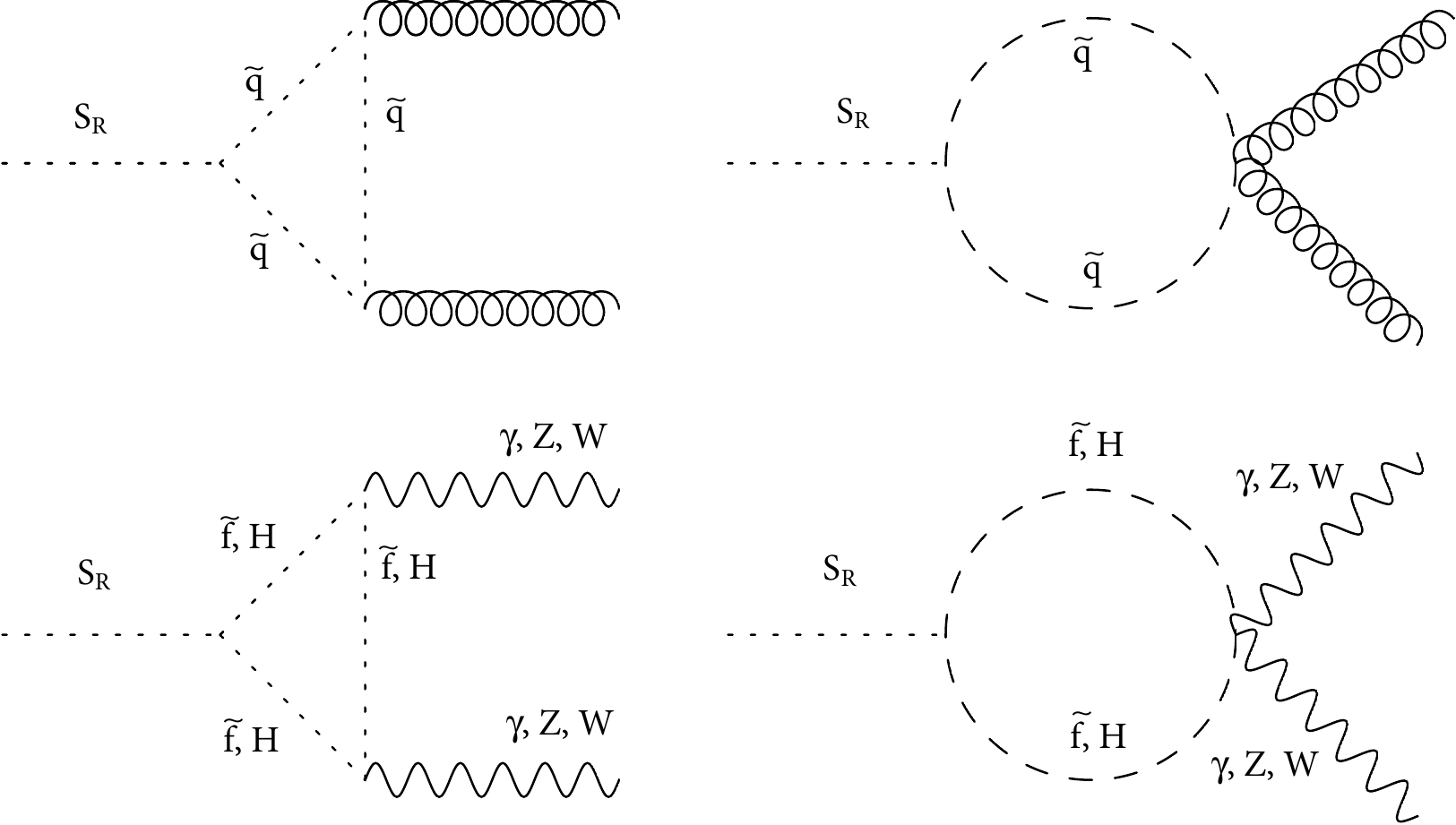}
\caption{One loop diagrams contributing to singlet coupling to pairs of gluons and electroweak gauge bosons.}
\label{fig:DecayLoop}
\end{figure}

\noindent There is also a coupling to the Higgs fields which we further discuss below.  We see that the real part of the scalar field will couple to pairs of SM gauge bosons through loops of sfermions.  The couplings will be proportional to the hypercharge of each field and the square of the Dirac mass.  The relevant diagrams can be seen in Fig. \ref{fig:DecayLoop}.  It should be noted that couplings from superpotential terms induce additional loop contributions to the decay of the singlet.  For example, in the $\slashed{\mu}$SSM \cite{Nelson:2002ca} or the MRSSM \cite{Kribs:2007ac} one would expect additional diagrams with charged Higgsinos running in the loops.  However, in what follows we have chosen not to fully specify the Higgs potential and we expect that these additional contributions will not qualitatively change our results as the charged Higgsinos are expected to be heavy.

We will see below that in order for a $750$ GeV sbino to reproduce the observed excess, we will need a Dirac gaugino mass on the order of 10 GeV.  Since the coupling in Eq. \ref{eq:cubic} is proportional to the Dirac mass, one may be concerned that such a large trilinear coupling will destabalize the electroweak breaking vacuum \cite{Carena:2012mw}.  However, the vacuum expectation value for S is typically 10 -100 GeV or less, leading to a diagonal mass contribution for the sfermions of $q g_Y m_D \sim 100 - 1000$ GeV.  This possible negative diagonal mass contribution is easily 
counteracted by positive soft mass contributions to the sfermions and thus we expect no tacyonic sfermion masses. 

To compute each contribution to the diphoton diagram, we must add up all fields which have hypercharge.  These fields include three generation of squarks and sleptons, $Q_L$,  $\overline{u}_R$,  $\overline{d}_R$ and $L_L$,  $\overline{l}_R$ plus the Higgs doublets, $H_u$ and $H_d$.  However, only the three generations of squarks contribute to the digluon channel.  Thus we expect to complete our operators by calculating loops of the roughly TeV scale scalar ``messengers".  We may control the ratio of the digluon channel to that of the of electroweak gauge boson channel by varying the mass scale of the squarks with respect to the sleptons.  Making the squarks much heavier than the sleptons will reduce the digluon coupling while maintaining or enhancing the coupling to pairs of electroweak gauge bosons.  In the limit of mass degenerate left and right handed states, the coupling of the singlet to dibosons will vanish due to a diagramatic cancelation.  This is equivalent to the cancelation which occurs in the coupling of the sgluon (color octet adjoint) to pairs of gluons as noted in reference \cite{Plehn:2008ae,Choi:2009ue}.  Thus, the strength of the effective operators coupling the sbino to gauge bosons may be dialed by changing the mass differences between left and right handed states. We also note that by varying the masses of left handed to right handed particles in general, we may control the ratio of the coupling of S to the U(1) and SU(2) gauge bosons.  In this way we may set the scales of our operators in Eq. \ref{eq:StoGaugeB}.

In general models of Dirac gauginos have quite a flexible spectrum.  In the simplest models, the mass ratio of the gauginos and sparticles differ by the square root of a loop factor.  However the mass splittings between gauginos and scalars can in principle be arbitrary.   In many completions of Dirac gaugino models, the physics which sets the scalar adjoint masses may also effect the SUSY spectrum; gauginos may have some mix of Dirac and Majorana masses, models may have extra R symmetric gauge mediated contributions, for example see  \cite{Fox:2002bu,Carpenter:2010as}.  In addition, SUSY models with Dirac gauginos are generally less  constrained by LHC searches than other SUSY scenarios \cite{Kribs:2012gx}.  The sparticle spectrum and decay chains are quite different in Supersoft scenarios than in other SUSY models, and mass constraints on squarks and sleptons can be weaked considerably.

\subsection{Tree Level Contributions to the Singlet Width }
From arguments above, we also note that the singlet state S couples to the Higgs fields.  In particular we the real part of the singlet couples to both $H_u$ and $H_d$ through the hypercharge D-term.
\be
V=  \frac{1}{2} g_Y m_D S_R (h_u {h_u}^{\dagger}-h_d {h_d}^{\dagger})
\ee

\noindent While the $H_d$ field presumably comprises the greater part of the heavy CP even Higgs state, the light Higgs $h_0$ is mostly $H_u$.  We thus expect that the singlet will have tree level decays into the light Higgs boson.  Since the scale $m_D$ is fairly large, we may expect an intermediate size width for the singlet driven by Higgs decays.  The trilinear coupling is proportional to $g_Y m_D \sim$ TeV, and the width is by

\be
\Gamma_{h}=\frac{(g_Y m_D)^2}{16 \pi m_S}  (\text{cos}^2\alpha-\text{sin}^2\alpha)^2\sqrt{1-4m_h^2/m_S^2}.
\ee

\begin{figure}[tbh]
\centering
\includegraphics[trim={0cm, 13cm, 3cm, 9cm},scale=.6]{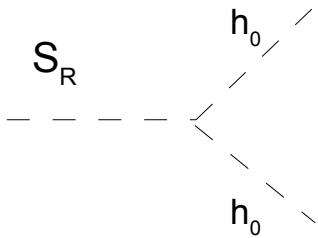}
\caption{SM singlet coupling to pairs of the lightest Higgs.}
\label{eq:SDecayToH}
\end{figure}

\noindent where $m_S$ is the mass of the singlet, $m_D$ is the Dirac bino mass and $\alpha$ is the Higgs mixing angle.  As a demonstrative partial width we may pick the Dirac bino mass to be 10 TeV, the singlet mass of $750$ GeV, and we choose to be in the decoupling limit with tan$\beta\sim$ 10 assuming minimal mixing with the singlet state.  With these numbers we may achieve a Higgs width of approximately 50 GeV.

Obtaining a width in the 10s of GeV range is a challenge for many models.  However, this intermediate sized width is achieved in this model because of the special coupling of the scalar to the Higgs states is proportional to a rather large mass scale, that is $g_Y m_D$.  In principle the singlet state, along with the SU(2) triplet may mix with the Higgs. The physics of such a Higgs sector in Dirac gaugino models has been studied, for example in references \cite{Carpenter:2015mna,Benakli:2012cy,Diessner:2014ksa,Bertuzzo:2014bwa}. In principle there may be regions of allowed parameter space in which the lightest Higgs contains some singlet admixture, this will effect the singlet width but may have effects on the light Higgs observables. The singlet may also couple to pairs of the heavy Higgses, however the tree level decay may be easily suppressed if the mass of the heavy states is more than half of the singlet mass.  Other Dirac gaugino models include $\mu$-less MSSM \cite{Nelson:2002ca} models in which the scalar is given a tree level coupling to $H_u$ and $H_d$ and the MRSSM \cite{Kribs:2007ac} which introduces two additional doublets, the R-Higgses.  This would result in a tree level coupling of the singlet to Higgsino (or R-Higgsino) pairs.  We will not consider the case here, however this might open another avenue for a decay of the singlet which would provide a sizable width into invisible or highly mass degenerate states.



Many models predict a large enhancement of the dihiggs production rate through resonant effects.  Current limits on dihiggs resonant production for a state of mass 750 GeV are tightest in the 4b channel \cite{Chen:2014ask,Aad:2014yja,Batell:2015koa}.  ATLAS has placed limits on the total production cross section times branching fraction in this channel at 42 fb at 8 TeV.  An extrapolation, then, for the limit on total rate into this channel at 13 TeV is approximately 200 fb \cite{Knapen:2015dap}.  As we expect most of our total singlet decay width to be taken up by dihiggs production, the total singlet production cross section will largely be limited by this constraint.

\section{Production and Decay}





We now discuss loop level couplings of the singlet field to pairs of electroweak gauge bosons.  As stated above the loop level coupling of the singlet to gluons is mediated by squark loops.
The value of the effective coupling of the singlet S to pairs of gluons is given by
\be
\frac{g_Y g_s^2}{16 \pi^2}\frac{m_D}{m_S^2}N_c (\Sigma q_{Q_L} C(0,0,m_s^2,m_{Q_L},m_{Q_L},m_{Q_L} ) + \Sigma q_{Q_R}  C(0,0,m_s^2,m_{Q_R},m_{Q_R},m_{Q_R}))
\ee
Here $C$ is the dimensionless Passerino-Veltman form factor,  $m_{Q_L}$ and  $m_{Q_R}$ are the masses of the left and right handed species of squarks, and the sums are taken over each species of left or right handed squark. The hypercharge of each species is given by $q_{Q_i}$ and we see that within each generation of squarks there will be cancelations due to sign differences in hypercharge.  If all squarks are to contribute to the loop we must sum over 3 generations of up and down type left and right handed squarks.

The value of the effective coupling of the singlet S to the U(1) gauge boson is given by

\be
\frac{g_Y^3 }{16 \pi^2}\frac{m_D}{m_S^2} ( \Sigma N_c   q_{L}^3 C(0,0,m_s^2,m_{L},m_{L},m_{L} ) + \Sigma N_c q_{R}^3  C(0,0,m_s^2,m_{R},m_{R},m_{R}))
\ee
where the sums are taken over every state with hypercharge, not just over squarks.  Here $ q_{i}$ denotes the hypercharge of the field in the loop. Many more particles contribute to this coupling than to the effective coupling of the singlet to gluons.  In principle the value of these two effective couplings depends on the hypercharges and masses of the light states that contribute to the loop. We may choose various masses for the left and right handed, and up and down type squarks and sleptons, and a complete spectrum will be given by the parameters of the high energy theory.  Here however we may simplify the theory by considering some particles to be arbitrarily high in mass.  There are several options for this `simplified model' of Supersoft SUSY.  One choice might be to consider all squarks except the lightest, presumably the right handed stop, to be very massive.

For the sake of simplicity we will consider a slightly different simplified model.  We will send the soft masses of all of the left handed states very high, effectively decoupling them from the theory.  This will serve as an existence proof that we may find points in parameter space which fit the excess.  We will assume that the only sfermions to have small soft masses will be the right-handed squarks and right-handed sleptons.  We will also set the masses of all 3 generations of up and down right handed  squarks to be equal. We will will consider that masses of all three generations of sleptons are equal.  In this way the couplings of the singlet and SU(3) and U(1) gauge boson are of comparable order.

\begin{figure}[H]
\centering
\includegraphics[scale=.8]{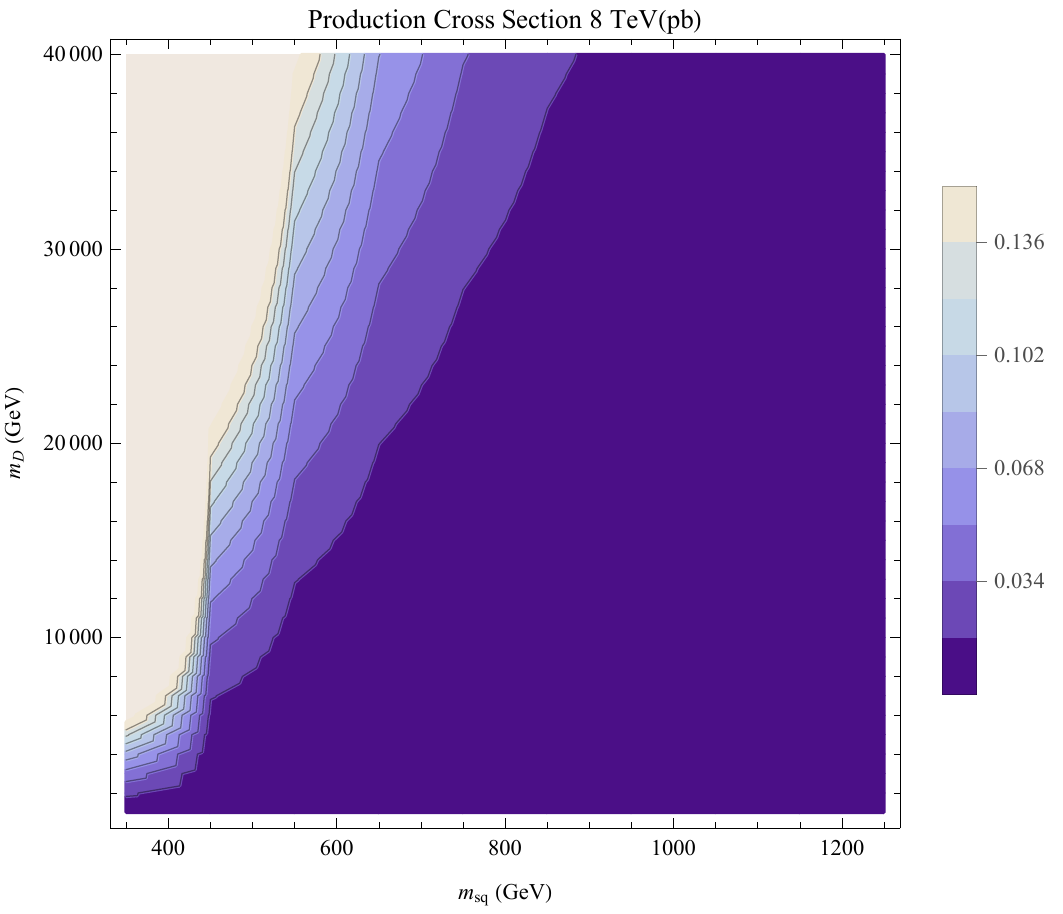}
\includegraphics[scale=.8]{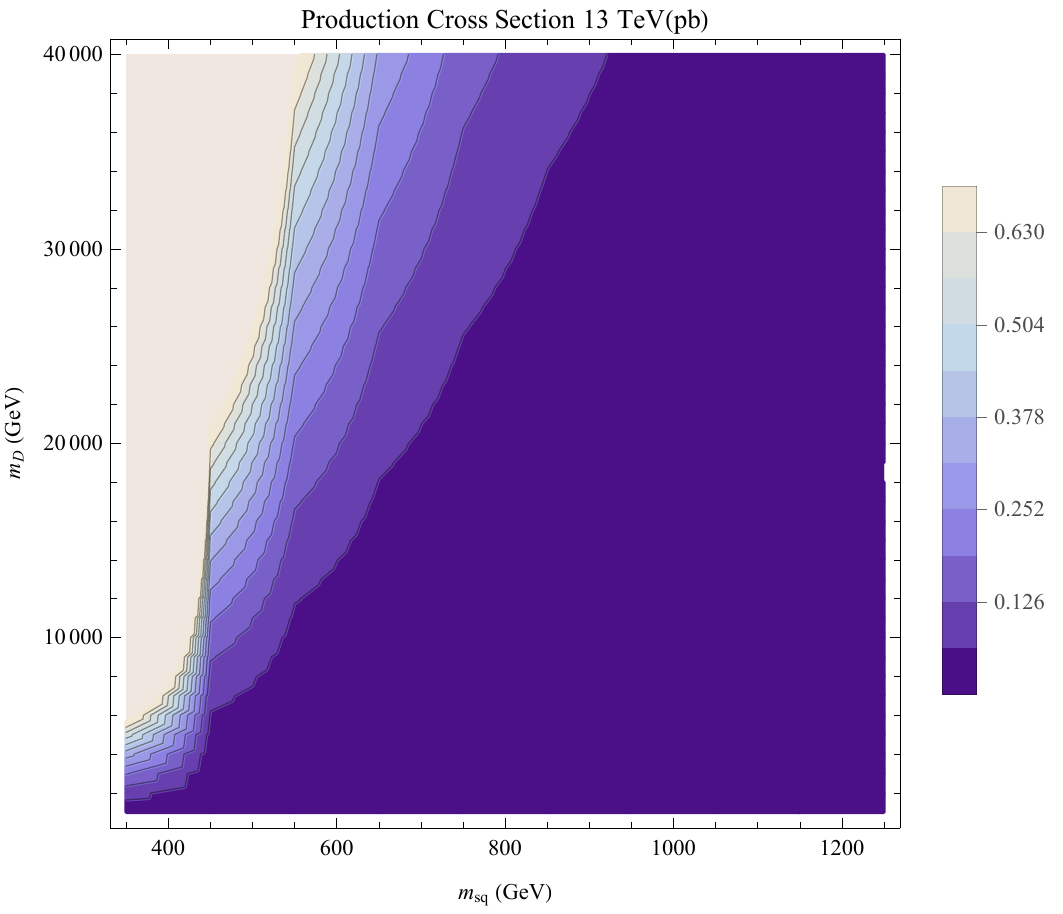}
\caption{$gg\rightarrow S$ rate in pb for squark mass vs Dirac mass plane in p-p collisions at 8 TeV and 13 TeV.}
\label{fig:xsectCon8}
\end{figure}

With the singlet coupling to gluons above we may now calculate gluon fusion production rates for the singlet.  In order to calculate the production cross section of the process $gg\rightarrow S$ in proton-proton collisions, we have implemented a our model into Feynrules \cite{Alloul:2013bka} inputting right handed up type squarks with D-term coupling to the sbino singlet field. This model was renormalized using the the NLOCT \cite{Degrande:2014vpa} package and imported into Madgraph5@NLO \cite{Alwall:2014hca} to calculate cross sections. The singlet mass was set to 750 GeV and cross sections were calculated. Below, in Fig. \ref{fig:xsectCon8}, we show contour plots of the total gluon fusion in the process $pp\rightarrow S$ in the $m_{\tilde{q}}$ vs.  $m_{D}$ plane at 13 and 8 TeV.

\noindent Again, the total production cross section will depend on how many light squarks are allowed to run in the loop, and what their masses and hypercharges are.  The total production rate via gluon fusion  increases with the square of $m_D$.  The cross section also increases dramatically as the squarks are made lighter.

\begin{figure}[tbh]
\centering
\includegraphics[scale=.8]{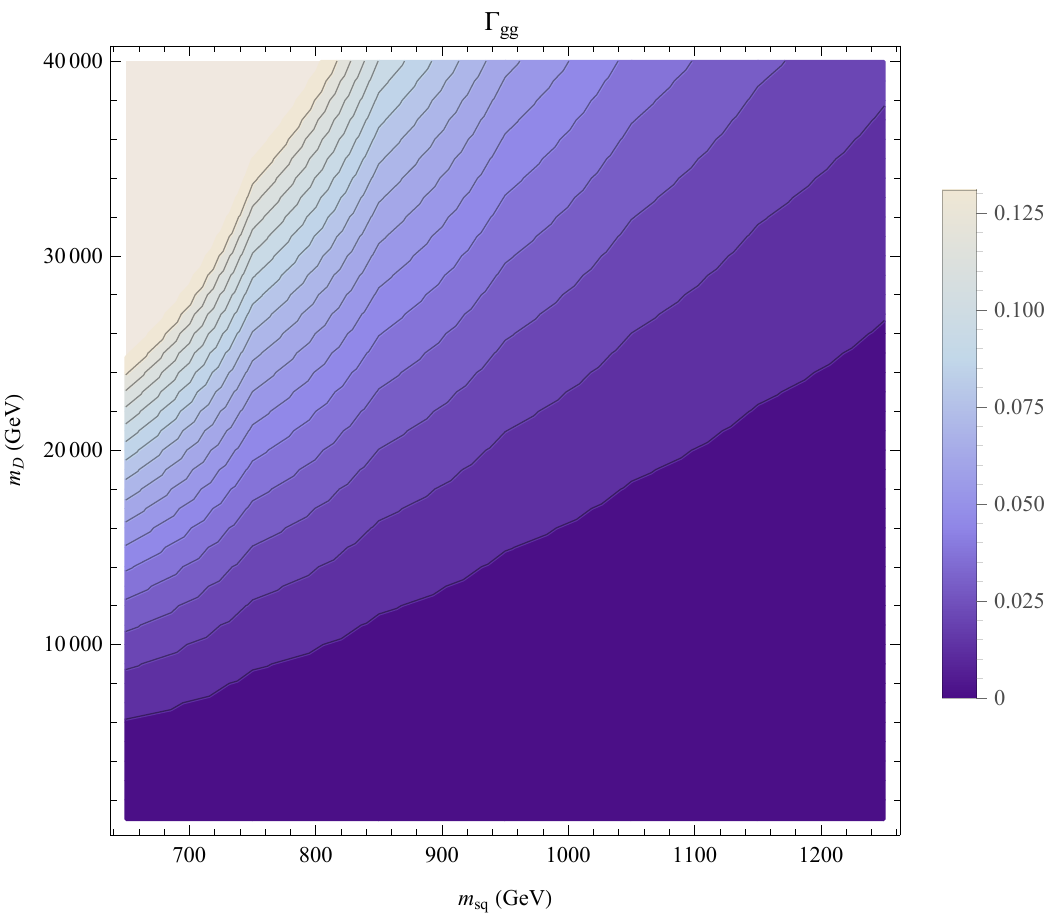}
\caption{Partial width (GeV) of S to gluon gluon in $m_D$-$m_s$ plane.}
\label{fig:ggWidth}
\end{figure}

We may also compute the partial decay width, $\Gamma_{gg}$, of the singlet state into digluon pairs.  This depends on the chosen value of $m_D$ as well as the value of the squark masses.  In Fig. \ref{fig:ggWidth} below we show a contour plot of the this width in the $m_D-m_{\tilde{q}}$ plane.

The ratio of the S width to photons and to gluons depends on the loop form factors, but also on the admixture of fields contributing to the photon loop vs the gluon loop.   In our simple model with only right handed squarks and sleptons contributing to loops, we can express the ratio of partial widths of the singlet into photons over partial width into gluons as

\be
\Gamma_{\gamma \gamma}/\Gamma_{gg}= \frac{c_w^4 g_Y^4}{N_c^2 g_3^4}\left( \frac{\Sigma q_{l_R}^3  C(0,0,m_s^2,m_{l_R},m_{l_R},m_{l_R})+\Sigma N_c q_{Q_R}^3
  C(0,0,m_s^2,m_{Q_R},m_{q_R},m_{q_R})}{\Sigma q_{Q_R}  C(0,0,m_s^2,m_{Q_R},m_{Q_R},m_{Q_R}) }\right)^{2}.
\ee

\begin{figure}[tbh]
\centering
\includegraphics[scale=.8]{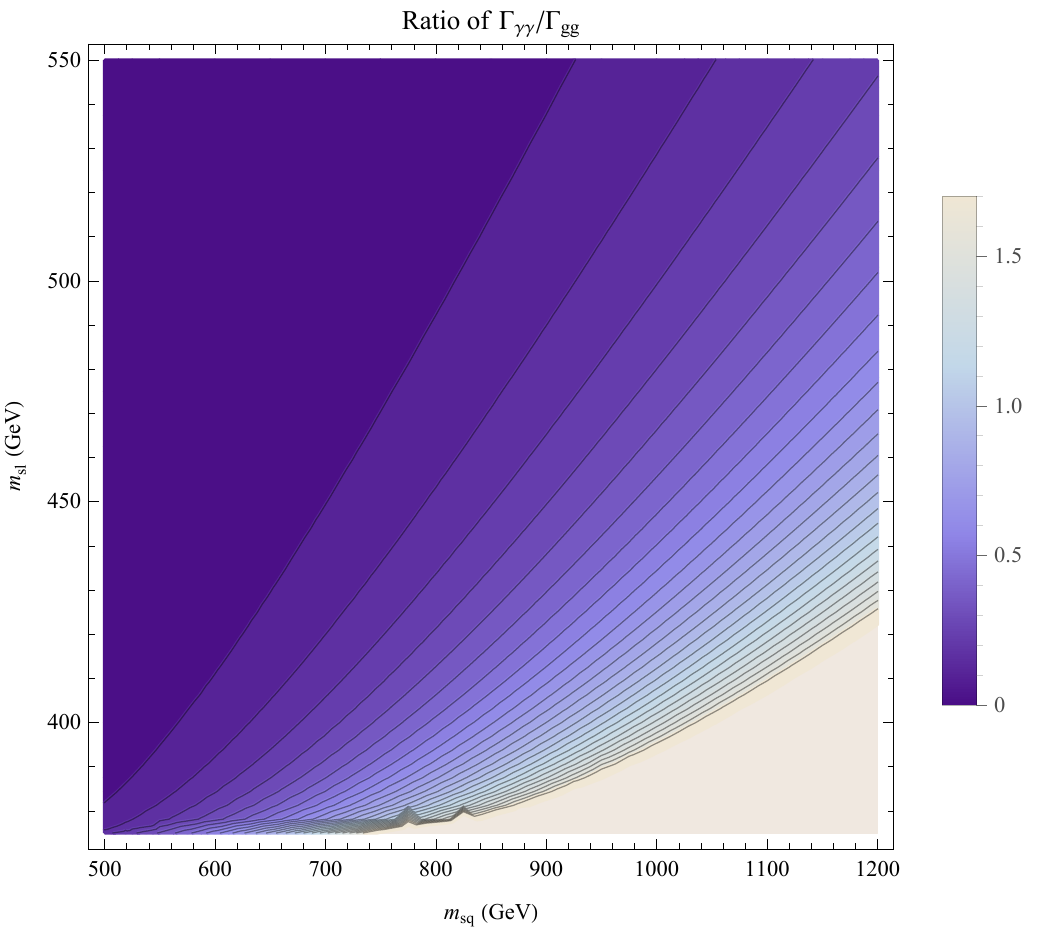}
\includegraphics[scale=.8]{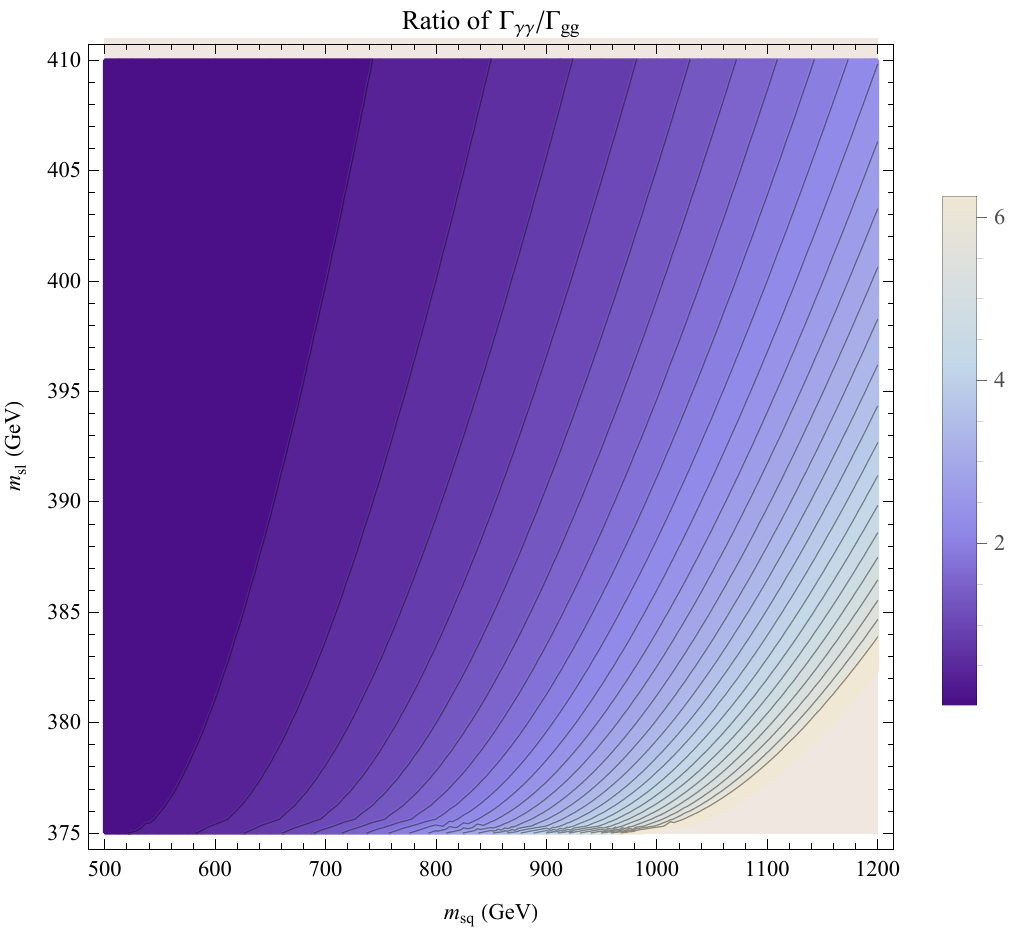}
\caption{Ratio of diphoton to digluon partial widths in the squark mass vs. slepton mass plane.  The right plot is a zoomed-in image of the plot on the left.}
\label{fig:RatioCon}
\end{figure}

\noindent The form factors are quite sensitive to the ratio of the squark and slepton masses and the ratio is independent of the Dirac mass.  We see that we may vary the ratio of partial widths to gluons and photons by varying the squark and slepton masses.  In particular, in the regime that the squarks are heavier than the sleptons, we find that the partial width to photons may be made appreciable. Below we have created a contour plot of the ratio of decay widths of the singlet to gluons and photons over the squark, slepton mass plane Fig. \ref{fig:RatioCon}.

There are many points in the overall parameter space that give a total diphoton rate in the 5-10 fb range at the 13 TeV run of LHC.  These points generally require $m_D$ to be in the 10 TeV range, with squarks masses in the 700-1000 GeV range.  In order to boost the diphoton rate we must pick slepton masses lighter than the squark mass but still heavier than half the mass of the singlet field.  An example point has values $(m_D, m_{\tilde{q}},m_{\tilde{l}} )$ of  $(50~\text{TeV}, 650~ \text{GeV},380~\text{GeV} )$ with a diphoton rate $\sigma_{\gamma \gamma}$ of 5 fb.  We also note at this point in parameter space the partial width to gluons, $\Gamma_{gg}$, is 0.411 GeV, while the partial width to diphotons $\Gamma_{\gamma \gamma}$, is 0.411 GeV.  We have considered that the total decay width of the singlet is 30 GeV.  The branching fraction into photons at this point is roughly 1.4$\%$.  At this point in parameter space, the total production cross section is 82 fb at 8 TeV which is within bounds for dijet and dihiggs production.  The branching ratios and production cross sections will change if we change the admixture of light squarks and sleptons.  For example we expect a reduced overall production cross section in the case that the only light squark is a right handed squark.  However, the ratio of diphoton to digluon events is also quite different.

It seems that this point in parameter space can be roughly in accordance with 8 TeV constraints in alternate singlet channels. However we do note that if the singlet decays into two photons, it will also decay into other pairs of electroweak gauge bosons with fixed ratios as dictated by gauge invarience. The coefficients for the effective operators coupling the singlet to diboson pairs is derived directly from the effective superpotential terms in Eq. \ref{eq:StoGaugeB} and, after accounting for $SU(2)\times U(1)$ breaking, are given by
\bea
g_{\gamma \gamma}&=& \frac{c_w^2}{\Lambda_1}+\frac{s_w^2}{\Lambda_2}  \\
g_{Z \gamma}&=& c_w s_w(\frac{1}{\Lambda_2}-\frac{1}{\Lambda_1}) \nonumber \\
g_{ZZ}&=&\frac{s_w^2}{\Lambda_1}+\frac{c_w^2}{\Lambda_2} \nonumber \\
g_{WW}&=& \frac{1}{\Lambda_2} \nonumber
\eea

\noindent where $\Lambda$ is the effective operator cut-off scale. For our scenario we might expect the $ZZ$ and $\gamma Z$ resonances at a slightly smaller but same order of production rate. Thus we expect dihiggs and diweak boson signals to follow close on the heel of this excess.

\section{Conclusions}
We have proposed that the LHC diphoton excess may be explained minimally in a model with Dirac gauginos. The signal follows from production and decay of the sbino, the real component of the scalar partner of the field which gives Dirac mass to the bino.  This particle is a scalar SM singlet and has a large tree level decay width to pairs of Higgs bosons, and possibly also to Higgsinos.  The sbino couples to pairs of dibosons through loops of squarks and sleptons.

As a simple viability proof we have calculated the production cross section and decays of the sbino in a very minimal model in which we consider only the loop contributions of up type right handed squarks and right handed sleptons.  We find the the sbino-gluon-gluon coupling, and hence the gluon fusion production cross section of the sbino, is highly sensitive to the squark masses running in the squark loop.  We have produced production cross sections for the full one loop computation. Due to large tree level decays into Higgs, the total branching fraction of the sbino into dibosons is percent level.  However, we find that when the sleptons are lighter than the squarks the partial width of the sbino into photons may be appreciable.

In general the most constraining feature in our model is a large rate in the resonant dihiggs channel.  The model generally predicts several hundred fb's of dihiggs production at 13 TeV.  We have shown this is within the bounds of LHC exclusion limits at 8 TeV.  It is possible that with further studies of  parameter space, this rate may be suppressed. For example, by opening up decays to light Higgsinos which would appear to searches as extremely mass degenerate particles.  Another option may be exploring low tan$\beta$ regions.  This would require further study in the Higgs sectors of the model to determine viability.
We find that with a small branching fraction into gluons, we do not run aground of the 8 TeV dijet search.

It is a topic of further study to determine how a less simplistic mass spectrum would effect the diphoton rate.  Spectra in Dirac gaugino models can become quite complex. The Dirac masses are independent and yield `supersoft' mass contributions to the squarks and sleptons. The gauge mediation mechanism can be mixed with other mediation mechanisms, for example anomaly mediation, which could produce interesting effects like mass splitting between the third generation and others see for example \cite{Carpenter:2005tz}. In addition mechanisms which set the sbino, swino, and sgluon masses can effect the sfermion masses.

As many have already pointed out, due to gauge invariance, if our state produced a diphoton signal, it must also be seen in the $ZZ$ and $Z\gamma$ and possibly WW channels.  These production rates are only a factor of a few less than the diphoton rate, and thus must be observed soon if the resonance is to stay.  As a final note, we mention that Dirac gaugino models also contain an scalar SU(2) adjoint. Loops similar to those we discussed couple the neutral component of the SU(2) adjoint to gluons and photons.  In the case of gluon coupling, the loops come with powers of Higgs insertions to preserve SU(2) quantum numbers.  This state also couples to electroweak boson pairs through loops of sleptons and it may therefore be possible to explain the excess as a swino.

\section{Acknowledgements}
This work was made possible with funds from DOE grant DE-SC0013529.

\end{document}